\documentstyle [twoside,12pt]{article}
\newcommand{\nc}{\newcommand}
\nc{\la}{\lambda} \nc{\La}{\Lambda}  \nc{\al}{\alpha}
\nc{\te}{\theta}  \nc{\be}{\beta}
\nc{\ga}{\gamma}  \nc{\Ga}{\Gamma}
\nc{\de}{\delta}  \nc{\De}{\Delta}
\nc{\si}{\sigma}  \nc{\ka}{\kappa}
\nc{\om}{\omega}  \nc{\Om}{\Omega}
\nc{\nf}{\infty}   \nc{\nl}{\newline}
\nc{\ra}{\longrightarrow}
\nc{\beq}{\begin{equation}}
\nc{\eeq}{\end{equation}}
\nc{\beqa}{\begin{eqnarray}}  \nc{\dst}{\displaystyle}
\nc{\eeqa}{\end{eqnarray}} \nc{\nnb}{\nonumber}
\title{\bf Einstein\,-Weyl structures and Bianchi metrics}
\author{Guy Bonneau\thanks {\noindent Laboratoire de Physique Th\'eorique et
des
Hautes Energies,
 Unit\'e associ\'ee au CNRS URA 280,~Universit\'e Paris 7,
 2 Place Jussieu, 75251 Paris Cedex 05. bonneau@lpthe.jussieu.fr}}
\begin{document}
\maketitle
\begin{abstract}
\noindent We analyse in a systematic way the (non-)compact
 four dimensional Einstein-Weyl spaces equipped with a Bianchi metric. We show
that Einstein-Weyl structures with a Class A Bianchi metric have a conformal
scalar curvature of constant sign on the manifold. Moreover, we prove that most
of them are conformally Einstein or conformally K\"ahler ; in the non-exact
Einstein-Weyl case with a Bianchi metric of the type $VII_0\,,\ VIII$ or $IX$,
we
show that the distance may be taken in a diagonal form and we obtain its
explicit 4-parameters expression. This extends our previous analysis, limited
to
the diagonal, K\"ahler Bianchi $IX$ case.
\end{abstract}

\vfill {\bf PAR/LPTHE/98-06/gr-qc/9803009}\hfill  February 1998
\newpage

\section{Introduction} In the last years, Einstein-Weyl geometry has raised
some
interest, in particular when in a recent paper, Tod
\cite{Tod96} exhibits the relationship between a particular Einstein-Weyl
geometry
without torsion (the four-dimensional self-dual Einstein-Weyl
 geometry studied by Pedersen and Swann \cite{PS93}) and local heterotic
geometry (i.e.  the Riemannian geometry with torsion and three complex
structures, associated with  (4,0) supersymmetric non-linear $\si$ models
 \cite{{HullWitten85},{xxy},{Delduc}}).

To extend these ideas to other situations, we analysed in a first step
\cite{bonneau1}(hereafter referred to as [GB]) Einstein-Weyl equations in the
subclass of diagonal K\"ahler Bianchi $IX$ metrics (in the standard
classification
\cite{DS95,Tod95}).  In the present work, we study (non-)compact 4-dimensional
Einstein-Weyl structures (for recent reviews see refs.
\cite{{PS93},{Madsen-a}})  on cohomogenity-one  manifolds with a 3 dimensional
group of isometries transitive on codimension-one surfaces, {\it i.e.}, in the
general relativity terminology,  Bianchi metrics, and neither require a
diagonal
metric nor the K\"ahler property ; we however obtain interesting results for
any
(class A) Bianchi metrics.

Let us recall that, in the compact case, on general grounds, strong results on
Einstein-Weyl structures have been known for some time :
\begin{itemize}
\item There exits a unique metric in a given conformal class [g] such that the
Weyl form is co-closed \cite{Gauduchon}\ , $$ \nabla_{\mu}\ga^{\mu} = 0\ .$$
One
then speaks of the "Gauduchon's gauge" and of a "Gauduchon's metric".
\item The analysis of Einstein-Weyl equations in this gauge gives two essential
results :
\begin{itemize}
\item  The dual of the Weyl form $\ga$ is a Killing vector \cite{Tod92}:
$$\nabla_{(\mu}\ga_{\nu)} = 0\ ,$$
\item Four dimensional Einstein-Weyl spaces have a constant conformal scalar
curvature \cite{PDSRheine}:
$$\nabla_{\mu} S^D = -\frac{n(n-4)}{4}\nabla_{\mu} (\ga_{\nu} \ga^{\nu})\ .$$
\end{itemize}
\end{itemize}  The paper is organised as follows :  in the next Section, we
recall
the classification of Bianchi metrics and the expressions of geometrical
objects,
separating the 4-dimensional metric g into a "time part" and a 3-dimensional
homogeneous one. Focussing ourselves on Class A Bianchi metrics, we exhibit a
specific Gauduchon's gauge and show how, {\it in the diagonal case} the
Einstein-Weyl equations simplify and ensure that the dual
$\ga_{\mu}$ of the Weyl one-form $\ga$ is a Killing vector, as in the compact
case, and that the metric is either conformally Einstein or conformally
K\"ahler.
In particular, this proves that four-dimensional Einstein-Weyl spaces equipped
with a diagonal Bianchi $IX$ metrics are necessarily conformally  K\"ahler,
{\it
i.e.} that our previous solution [GB] is the general one, up to a conformal
transformation. \nl \noindent
In Section 3, we show that for {\it all class A Bianchi metrics},
there exits a simple Gauduchon's gauge such that the conformal scalar curvature
is
constant on the manifold and the dual
$\ga_{\mu}$ of the Weyl one-form $\ga$ satisfies
$D^{\mu}\left(\nabla_{(\mu}\ga_{\nu)}\right) = 0\ ,$  where $D$ denotes the
covariant derivative with respect to the Weyl connection $\ga\,.$ Using these
results, we prove that for Bianchi $VI_0\,,\ VII_0\,,\ VIII\,,$ and $IX$, the
most general
solution of Einstein-Weyl constraints is the same as the one in the diagonal
case, {\it i.e.} in the non-conformally Einstein cases, the K\"ahler one of
previous subsection, up to a conformal
transformation. Finally, we also prove that the only self-dual Einstein-Weyl
structures
are the Bianchi $IX$ ones of Madsen \cite{{Madsen-a},{Madsen-b}}.

\section{Bianchi metrics and Einstein-Weyl structures.}
\subsection{The geometrical setting}
\begin{itemize}
\item A Weyl space \cite{PS93} is  a conformal manifold with a torsion-free
connection D and a one-form $\ga$ such that for each representative metric g in
a
conformal class [g],
\beq\label{a1}
 D_{\mu} g_{\nu \rho} = \ga_{\mu} g_{\nu \rho}\ .
\eeq  A different choice of representative metric : $g\ \ra\
\tilde{g} = e^f g$ is accompanied by a change in $\ga\ : \ga\
\ra\ \tilde{\ga} = \ga + df\ .$  Conversely, if the one-form $\ga$ is exact,
the
metric g is conformally equivalent to a Riemannian metric $\tilde{g}$ : $
D_{\mu}\tilde{ g}_{\nu \rho} = 0.$ In that case, we shall speak of an {\it
exact}
Weyl structure.

\item On the other hand, Bianchi metrics are real four-dimensional metrics with
a
three-dimensional isometry group, transitive on 3-surfaces. Their
classification
was done by Bianchi in 1897 according to the Lie algebras of their isometry
group,
{\it i.e.} according to the Lie algebra structure constants
$C^i_{jk}\ ,(i,\,j,\, k\ =1,\,2,\,3\,)$ ; on general grounds, these ones may be
decomposed into two parts \cite{Landau}:
\beq\label{2-10} C^i_{jk} = n^{il}\epsilon_{jkl} + a_l[\de^i_j\de^l_k -
\de^i_k\de^l_j]
\eeq where the symmetric 3$\times$3 tensor $n^{il}$ may be reduced to a
diagonal matrix
with entries 0,1 or -1 and the vector $a_l$ satisfies
$$n^{il}a_l = 0\,.$$ This splits  Bianchi metrics into two classes : class A in
which the vector
$a_l$ is zero, and class B in which it has one non vanishing component, say
$a_1$.

\item An invariant Weyl struture may then be written as :
\beqa\label{2-11} ds^2  = g_{\mu \nu}dx^{\mu}dx^{\nu} = dT^2 +
h_{ij}(T)\si^i\si^j\
\ & ; & \ \ \ga  =  \ga_0(T)dT + \ga_i(T)\si^i\ ,
\nnb \\ d\si^i =\frac{1}{2}C^i_{jk}\si^j\wedge\si^k\ \ i,j,k = 1,2,3\  & ; & \
\mu,\ \nu = (0,\al),\ (0,\be)\ \ ;\ \si^i = \si^i_{\al}dx^{\al} \\
\si^i_{\al}\si^j_{\be}h_{ij}(T)  =  g_{\al \be}\ & ; & \
\si_i^{\al}\si^j_{\al} = \de_i^j\ ,\ \si_i^{\al}\si^i_{\be} =
\de^{\al}_{\be}\ , \nnb
\eeqa where the three $\si^i$ are one-forms invariant under the isometries of
the
homogeneous 3-space, characterised by the aforementionned structure constants
$C^i_{jk}\,.$  Notice that there is no loss of generality in choosing the
metric
element $g_{00} = 1$ as this corresponds to a choice of "proper time " T, but
the
matrix $h_{ij}$ is {\it a priori} non-diagonal. On another hand, one might
always choose a representative in the conformal class [g] such that $\ga_0 (T)
\equiv 0\,.$

\noindent The Ricci tensor associated to the Weyl connection D is defined by :
\beq\label{a2}  [D_{\mu},\,D_{\nu}\,]v^{\rho} = {\cal R}^{(D)\rho}_{\ \ \ \
\la,\mu\nu}\ v^{\la}\ \ ,\ \ {\cal R}^{(D)}_{\mu\nu} = {\cal R}^{(D)\rho}_{\
\
\ \ \mu,\rho \nu}\ .
\eeq
${\cal R}^{(D)}_{\mu\nu}$ is related to  $R^{(\nabla)}_{\mu\nu}$, the Ricci
tensor
associated to the Levi-Civita connection [GB]:
\beq\label{a3} {\cal R}^{(D)}_{\mu\nu} =  R^{(\nabla)}_{\mu\nu} +
\frac{3}{2}\nabla_{\nu}\ga_{\mu} - \frac{1}{2}\nabla_{\mu}\ga_{\nu} +
\frac{1}{2}\ga_{\mu}\ga_{\nu} + \frac{1}{2} g_{\mu\nu} [\nabla_{\rho}\ga^{\rho}
-
\ga_{\rho}\ga^{\rho} ]\ .
\eeq
\end{itemize}
$R^{(\nabla)}_{\mu\nu}$ may be expressed as (for exemple see \cite{Landau}) :
\beqa\label{2-12} R^{(\nabla)}_{00} & = &
-\frac{1}{2}\frac{d}{dT}(\frac{h'}{h}) -
\frac{1}{4}K_i^j K_j^i\ \ ;\ K_i^j = \frac{dh_{ik}}{dT}h^{kj}\ \ ;\ h =
\det[h_{ij}]\ , h' = \frac{dh}{dT}\ , \nnb\\ R^{(\nabla)}_{0\al} & = &
\frac{1}{2}\si_{\al}^k [C^i_{jk} - \de^i_k C^m_{mj}] K_i^j\ ,\\
R^{(\nabla)}_{\al\be} & = & \si_{\al}^i\si_{\be}^j\left[ R^{(3)}_{ij} -
\frac{1}{2}\frac{dK_{ij}}{dT} + \frac{1}{2}K_i^k K_{kj} -
\frac{h'}{4h}K_{ij}\right]\ ;\ K_{ij} = K_i^k h_{kj} = \frac{dh_{ij}}{dT}\ ,\
{\rm
e.t.c...}\nnb
\eeqa where $ R^{(3)}_{ij}$, the 3-dimensional Ricci tensor associated to the
homogeneous space Levi-Civita connection, in the basis of the one-forms
$\si^i$, may be expressed as a function of the 3-metric
$h_{ij}$ and of the structure constants of the group
\cite{{Landau},{russes}}.

\noindent In the same way, the 4-dimensional Bianchi identity splits :
\beq\label{2-13a} C^i_{jk}R^{(3)j}_i + C^i_{ji}R^{(3)j}_k = 0\ \ ,\ k =1,2,3\ \
\ (\cite{Landau},\ equ.(116,26))
\eeq and  (see the
appendix) :
\beq\label{2-13b} h^{ij}\frac{d}{dT}R^{(3)}_{ij} \equiv
\frac{dR^{(3)}}{dT} + K_i^jR^{(3)i}_j = 2C^i_{ji}R_0^j\ \ ,\ {\rm with}\ R_0^j
=
h^{ji}\si^{\al}_iR^{(\nabla)}_{0\al}\ \ ,\ R^{(3)} =R^{(3)}_{ij}h^{ij}\ .
\eeq We do not find equation (\ref{2-13b}) in the standard textbooks on
gravity.

\subsection{The Gauduchon's gauges} We computed (using equations
(\ref{A-0},\ref{A-1} of the appendix) the components of the tensor
$\nabla_{(\mu}\ga_{\nu)}$ and find ;
\beqa\label{2-21}
\nabla_{0}\ga_0 & = & \frac{d\ga_0}{dT}\nnb \\
\nabla_{(0}\ga_{\al)} & = &
\frac{1}{2}\si_{\al}^i h_{ij}\frac{d\ga^j}{dT}\\
\nabla_{(\al}\ga_{\be)} & = & \frac{1}{2}\si_{(\al}^i\si_{\be)}^j  [\ga_0
K_{ij} +
2\ga^k h_{il}C^l_{kj}]\ , \nnb
\eeqa and, as a consequence,
\beq\label{2-22}
\nabla_{\mu}\ga^{\mu} = \frac{1}{\sqrt{h}}\frac{d}{dT}[\sqrt{h}\ga_0]
-C^i_{ij}\ga^j\ .
\eeq When $C^i_{ij} \equiv 2a_j = 0$, which corresponds to class A Bianchi
metrics, a special Gauduchon's gauge is obtained through the choice :
\beq\label{2-23}
\ga_0(T) \equiv 0\,.
\eeq In the compact case, the choice (\ref{2-23}) is the unique good one
(\cite{Madsen-a},Proposition 5.20).

\subsection{The Einstein-Weyl equations}

Einstein-Weyl spaces are defined by :
\beqa\label{a5}  {\cal R}^{(D)}_{(\mu\nu)}  & = & \La ' g_{\mu\nu}\
\ \Leftrightarrow \nnb \\
 R^{(\nabla)}_{\mu\nu} +  \nabla_{(\mu}\ga_{\nu)} +
\frac{1}{2}\ga_{\mu}\ga_{\nu} & = & \La\  g_{\mu\nu}\ \ ,\ \ \La  =
\La' - \frac{1}{2}[\nabla_{\la}\ga^{\la} - \ga_{\la}\ga^{\la}] \ .
\eeqa Note that for an exact Einstein-Weyl structure, $\ga = df$, the
representative metric is conformally Einstein. Note also that the conformal
scalar
curvature is related to the scalar curvature through:
\beq\label{a6}  S^{(D)} = g^{\mu\nu}{\cal R}^{(D)}_{\mu\nu}  = 4\La +
2[\nabla_{\la}\ga^{\la} - \ga_{\la}\ga^{\la}] = R^{(\nabla)} +
3[\nabla_{\la}\ga^{\la} -
\frac{1}{2}\ga_{\la}\ga^{\la}]\ .
\eeq

For Class A Bianchi metrics, in the special Gauduchon's gauge (\ref{2-23}),
Einstein-Weyl constraints (\ref{a5}) splits into :
\beq\label{3-1-a}
\La   =   -\frac{1}{2}\frac{d}{dT}(\frac{h'}{h}) -
\frac{1}{4}K_i^j K_j^i\ \ ,
\eeq
\beq\label{3-1-b} n^{ij}\epsilon_{jkl}K_i^k = - h_{li}\frac{d\ga^i}{dT}\ ,
\eeq
\beq\label{3-1-c}
\La h_{ij} =  R^{(3)}_{ij} -
\frac{1}{2}\frac{dK_{ij}}{dT} + \frac{1}{2}K_i^k K_{kj} -
\frac{h'}{4h}K_{ij} + \frac{1}{2}\ga_i\ga_j +
\frac{1}{2}\ga^k[h_{im}n^{mn}\epsilon_{nkj} + h_{jm}n^{mn}\epsilon_{nki}]\ .
\eeq

\subsection{Diagonal metrics and conformal K\"ahlerness}   Let us restrict
ourselves to the diagonal  Bianchi metrics, usually written as
\cite{{DS95},{Tod95}}:
\beq\label{2-30} ds^2 = \om_1\om_2\om_3 (dt)^2
+\frac{\om_2\om_3}{\om_1}(\si^1)^2
+
\frac{\om_1\om_3}{\om_2}(\si^2)^2 + \frac{\om_1\om_2}{\om_3}(\si^3)^2
\eeq Define $\al_i$ through :
\beq\label{2-301}
\frac{d\om_i}{dt} = \al_i\om_i + n^{ii}\om_j\om_k\ \ ,\ \ (i,j,k) =  {\rm
circ.\
perm.}\ (1,2,3)\ .
\eeq

\noindent In \cite{DS95}, Dancer and Strachan  gave the conditions on the
$\al_i$
under which the four dimensional diagonal Bianchi metric is K\"ahler, but not
Hyper-K\"ahler. These conditions  are :
\nl\noindent - Class A : two of the $\al_i$ have to be equal, the third one
vanishing ;
\nl\noindent - Class B : the three $\al_i$ have to be proportional to $\om_1$
and
to satisfy : $\al_1 = \al_2 + \al_3\,.$
\nl\noindent Under a conformal transformation preserving the cohomogeneity-one
character of a Bianchi metric : $\tilde{g} = \mu^2(T)g\,,$ these conditions are
easily converted into conditions for K\"ahlerness up to a conformal
transformation
:
\nl\noindent {\bf Lemma 1} : {\it A diagonal Bianchi metric (\ref{2-30}) is
conformal to a K\"ahler one iff. :}
\begin{itemize}
\item {\it Class A metric ($a_{i} = 0$) : two of the $\al_i$ are equal };
\item {\it Class B metric  ($a_i = a\de_{i1}$) : the following relations  hold}
:
$$ \frac{\al_1 - \al_3}{a\om_1} = \frac{a\om_1}{\al_2 - \al_1} = {\rm Cste}\
.$$
\end{itemize}

\noindent For a Class A diagonal Bianchi metric , equation (\ref{3-1-b}) leads
to
\beq\label{2-31}
\ga^i(T) = \Ga^i\ {\rm constant}\ ,
\eeq and equations (\ref{3-1-c}) wrote for $i\neq j$ :
\beqa\label{2-32}
\frac{1}{2}\ga_i\ga_i + \nabla_{(i}\ga_{j)} = 0\ \ & , & \ \ i\neq j\ \ \ \
\Leftrightarrow \nnb \\
\om_1\om_2\om_3\Ga^i\Ga^j = \Ga^k[n^{ii}\om_j^2 - n^{jj}\om_i^2]\ \ & , & \ \
(i,j,k) =  {\rm circ.\ perm.}\ (1,2,3)\ .
\eeqa  By inspection of the different possibilities for the $n^{ii}$
\cite{Landau}, it is readily shown that at least two of the $\Ga^i$ necessarily
vanish, with no other constraint for Bianchi $I$ and $II$ ; for Bianchi $VI_0$,
the three of them vanishing, the metric is necessarily conformally Einstein ;
for
Bianchi $VII_0$ and $VIII$ [$n^{11} = n^{22} = +1$] the third $\Ga^3$ is
constrained by
$$\Ga^3 [\om_1^2 - \om_2^2] = 0\ ,$$ \noindent then, either the metric is
conformally Einstein or, with $\om_1^2 = \om_2^2\,,$ the metric is conformally
K\"ahler (thanks to Lemma 1). For Bianchi $IX$ case, the same result holds, the
special direction being unfixed (it will be chosen in the same direction as for
Bianchi
$VII_0$ and $VIII$). A Corollary of this analysis is that in all Class A cases,
the dual of the one form
$\ga$ is a Killing vector.

In these three types of Bianchi metrics,
$$n^{11} = n^{22} = +1\ \ ,\ \ \ \Ga^1 = \Ga^2 = 0\ \ ,\ \ \om_1 =
\om_2 = \om \ \ ,\ \ \ \al_1 = \al_2 =\al \ \ ,$$
\noindent and the remaining equations (\ref{3-1-a},\ref{3-1-c}) wrote in the
vierbein basis corresponding to (\ref{2-30})( a comma indicates a derivative
with
respect to $t$):

\beqa\label{3-41}  \mbox{\scriptsize{(00)}}\ \ 2(\om)^2\om_3 \La &  = & -2\al'
-
\al_3' -(\al_3)^2 + 4\al_3\al + 2\al_3\om_3 + n^{33}\frac{(\om)^2}{\om_3}(2\al
-
\al_3)\nnb \\
\mbox{\scriptsize{(11,22)}}\ \ 2(\om)^2\om_3 \La & = &  - \al_3' -
n^{33}\frac{(\om)^2}{\om_3}(2\al -
\al_3)\  \\ \mbox{\scriptsize{(33)}}\ \ 2(\om)^2\om_3 \La & = & (\Ga^3)^2 \om^4
-2\al' + \al_3' - 2\al_3\om_3 + n^{33}\frac{(\om)^2}{\om_3}(2\al - \al_3)\nnb
\eeqa Consider  the function $u(t) = \frac{\al_3}{\om^2}\,.$ Its derivative is
readily obtained, using the difference of the {\scriptsize(00)} and
{\scriptsize(33)} equations (\ref{3-41}) :
$$ \frac{d u}{dt} = -\frac{1}{2}\om^2[(\Ga^3)^2 + u^2]\ \ \ <\ 0\,.$$
\noindent Then one can change the variable $t$ into $u$ and compute :
$$ \frac{d \om_3}{du} = - 2\frac{n^{33} + u\om_3}{ (\Ga^3)^2 + u^2}\ \
\,,$$ which integrates to :
\beq\label{3-42}
\om_3(u) = 2\frac{- n^{33}u + k}{ (\Ga^3)^2 + u^2}\,.
\eeq Then one gets :
\beq\label{3-43}
\al(u) = -2\frac{- n^{33}u + k}{ (\Ga^3)^2 + u^2} -
\frac{1}{4}[(\Ga^3)^2 + u^2]\frac{d\om^2}{ du}\ \ ,\ \ \al_3(u) = u\om^2(u)\,.
\eeq The difference of the {\scriptsize(11)} and {\scriptsize(33)} equations
(\ref{3-41}) then gives a second order linear differential equation on
$\om^2(u)$
:
\beqa\label{3-44}
\frac{d^2 \om^2}{d u^2} & + & \left[ \frac{6u}{(\Ga^3)^2 + u^2} -
\frac{2n^{33}}
{n^{33}u - k}\right]\frac{d \om^2}{d u}\nnb \\ & - & 4 \left[
\frac{(\Ga^3)^2}{((\Ga^3)^2 + u^2)^2} + \frac{k}{(n^{33}u - k)((\Ga^3)^2 +
u^2)}\right]\om^2 + \frac{8n^{33}}{((\Ga^3)^2 + u^2)^2} = 0\,.
\eeqa The solution is :
\beqa\label{3-45}
\om^2 & = & \frac{4}{(\Ga^3)^2 + u^2}\Om^2 \ ,\ {\rm with}\ \ \ \Om^2  =
n^{33}
+ \la_1[n^{33}(u^2 - (\Ga^3)^2) - 2ku] +\nnb \\ & + & \la_2\left[[n^{33}(u^2
-(\Ga^3)^2) - 2ku]\Ga^3\arctan(\frac{u}{\Ga^3}) +
(\Ga^3)^2[n^{33}u - 2k]\right]\ .
\eeqa  Equations (\ref{3-42},\ref{3-45}) and
\beq\label{3-46}
\frac{du}{dt} =  - 2\Om^2
\eeq give the distance \footnote{\ Of course, the 4 parameters $k,\ \Ga^3,\
\la_1,\ \la_2$ and the ``time" variable $u$ are constrained by positivity  :
$\Om^2 >0\ ,\
\ -n^{33}u +k >0\ .$} and Weyl form as functions of the new ``proper time" $u$
:
\beqa\label{3-47}  ds^2 & = & 2\frac{-n^{33}u + k}{ \Om^2((\Ga^3)^2 +
u^2)^2}(du)^2
 + 2\frac{-n^{33}u + k}{ (\Ga^3)^2 + u^2}[(\si^1)^2 + (\si^2)^2] +
2\frac{\Om^2}{-n^{33}u + k}(\si^3)^2 \ ,\nnb \\
\ga & = & \frac{2\Ga^3 \Om^2}{-n^{33}u + k}\si^3 \ \ .
\eeqa  Finaly, the conformal scalar curvature is the constant
\beq\label{3-48} S^D = 4\la_2(\Ga^3)^4\ .
\eeq  Under the conformal transformation $ \tilde{g} = [(\Ga^3)^2 + u^2] g/2$,
the
metric may be rewriten in the standard form (\ref{2-30}) with $$\tilde{\om}_1
=
\tilde{\om}_2 = \Om\sqrt{(\Ga^3)^2 + u^2)}\ ,\ \ \tilde{\om}_3  = -n^{33}u + k
\,,$$ the ``proper time" $\tilde{t}$ being given by
$$d\tilde{t} =-\frac{du}{\Om^2((\Ga^3)^2 + u^2)}\,.$$ Then,
$$\frac{d\tilde{\om}_3}{d\tilde{t}} - n^{33}\tilde{\om}_1\tilde{\om}_2 =
\tilde{\om}_3\tilde{\al}_3 = 0\ \ ,\tilde{\al}_1 = \tilde{\al}_2\ ,$$ ensuring
that the
metric
$\tilde{g}$ is K\"ahler.

Then we have proved the \nl\noindent {\bf Theorem 1}  : {\it The most general
(non-)compact  non-exact Einstein-Weyl structure with a diagonal Bianchi
$VII_0\,,VIII$ or $IX$ metric is conformal to a K\"ahler 4-parameters's one. In
particular, in the Bianchi $IX$ case, the K\"ahler metric is the one found in
[GB, equ.(27)].}
$$  $$

In the following Section, we shall consider non-diagonal Bianchi metrics
\footnote{\ When $\ga = 0$ (Einstein equations), and for Bianchi $VIII$ and
$IX$
metrics, it was shown in \cite{Tod95} that, thanks to (\ref{3-1-b}), the
looked-for Einstein metrics may be chosen to be diagonal. I thank Paul Tod for
a
clarifying discussion on that assertion.} , but still restrict ourselves to
Class
A ones, where the particular choice of Gauduchon's gauge (\ref{2-23}) will be
of great help.

\section{(Non-)compact Einstein-Weyl structures with class A Bianchi metrics.}

We first prove the

\noindent {\bf Lemma 2} : {\it In the special gauge $\ga_0 = 0$, Einstein-Weyl
structures with a Class A Bianchi metric have a constant conformal scalar
curvature $S^D$.}

Using $\nabla_{\mu}\ga^{\mu} = 0$, the conformal scalar curvature (\ref{a6})
writes
\beq\label{3-21} S^{(D)} = 4\La - 2\ga_i\ga^i\ .
\eeq Contracting equation (\ref{3-1-c}) with $K^{ij}$, using
(\ref{3-1-a},\ref{3-1-b}), leads to :
\beq\label{3-22} K^{ij} R^{(3)}_{ij} = \frac{d}{dT}\left[\frac{1}{4}K_i^j
K_{j}^{i} - (\frac{h'}{2h})^2 - \frac{1}{2}\ga_i\ga^i \right]\ .
\eeq Our Bianchi identity (\ref{2-13b}), with $C^i_{ij} = 2a_j = 0$, then gives
:
\beq\label{3-23} R^{(3)} + \frac{1}{4}K_i^j K_{j}^{i} - (\frac{h'}{2h})^2 -
\frac{1}{2}\ga_i\ga^i = \ {\rm Constant}\ .
\eeq Contracting now equation (\ref{3-1-c}) with $h^{ij}$, using (\ref{3-1-a}),
leads to :
\beq\label{3-24} R^{(3)} + \frac{3}{4}K_i^j K_{j}^{i} +
\frac{d}{dT}(\frac{h'}{h}) - (\frac{h'}{2h})^2 + \frac{1}{2}\ga_i\ga^i = 0
\eeq which, combined with (\ref{3-23},\ref{3-1-a}) gives
\beq\label{3-25} 2\La - \ga_i\ga^i \equiv S^{(D)}/2 = \ {\rm Constant\ \ \ \
Q.E.D.}
\eeq  We have the

\noindent {\bf Corollary 1} : {\it Einstein-Weyl structures with a Class A
Bianchi
metric have a conformal scalar curvature $S^{(D)}$ of constant sign on the
manifold.}

We may now prove the

\noindent {\bf Lemma 3} : {\it In any Gauduchon's gauge $\nabla_{\mu}\ga^{\mu}
=
0$, all Einstein-Weyl structures with a constant conformal scalar curvature
$S^D$
are such that
$D^{\mu}[\nabla_{(\mu}\ga_{\nu)}]$ vanishes. }

Acting with $\nabla^{\mu}$ on the Einstein-Weyl constraint (\ref{a5}) in the
Gauduchon gauge and using the four-dimensional Bianchi identity, the constant
value of $S^D \equiv R^{(\nabla)} -3/2\ga_{\mu}\ga^{\mu}$, one gets :
\beq\label{3-31}
\nabla^{\mu}[\nabla_{(\mu}\ga_{\nu)}] + \ga^{\mu}[\nabla_{(\mu}\ga_{\nu)}] =
-\frac{1}{4}\nabla_{\nu}S^D\ \ \ \ \ Q.E.D.
\eeq

\noindent Note that in the compact case, contraction of the previous identity
with
$\ga^{\nu}$, followed by an integration on the manifold, leads to the vanishing
of
$\nabla_{(\mu}\ga_{\nu)}$ \cite{Tod92}.
\begin{itemize}
\item Considering the $\nu = 0$ component of the previous equation, the
expression
of
$[\nabla_{(\mu}\ga_{\nu)}]$ given by (\ref{2-21}), and the formula (\ref{A-2})
given in the appendix, we obtain for any class A Bianchi metric :
\beq\label{3-32} h_{ij}\frac{d}{dT}(\frac{1}{2}\ga^i\ga^j) = 0\ .
\eeq

\item In the same way, considering the $\nu = \al$ component of the equation
(\ref{3-31}), and multiplying by $\si^{\al}_i$ gives after some manipulations
[using the expression of $\nabla_{\al}\si_{\be}^i $ given in the appendix
(\ref{A-1})]:
\beq\label{3-33}
\frac{d}{dT}(h_{ij}\frac{d}{dT}(\ga^j)) +
\frac{h'}{2h}(h_{ij}\frac{d}{dT}(\ga^j)) = C^j_{ik}\ga_j\ga^k +
X_{ij}[h_{mn}]\ga^j\ ,
\eeq where the 3$\times$3 symmetric matrix $X_{ij}[h_{mn}]$ is given by
$$X_{ij} =  h_{mn}h^{pq}C^m_{pi}C^n_{qj} + C^m_{ni}C^n_{mj}\,,$$ which may be
expressed for a Class A Bianchi metric as:
\beq\label{-3-331} X_{ij} = \frac{1}{\det h_{mn}}[Tr(hnhn)h_{ij} -(hnhnh)_{ij}]
+\epsilon_{ikr}\epsilon_{jls}n^{kl}n^{rs}\,.
\eeq
\end{itemize} The contraction of (\ref{3-33}) by $\ga^i$ and the use of
(\ref{3-32}), finally gives :
\beq\label{3-34}
\frac{d\ga^i}{dT}h_{ij}\frac{d\ga^j}{dT} + \ga^i X_{ij}\ga^j = 0\,.
\eeq
\noindent Then we have: \nl\noindent {\bf Lemma 4 } : {\it For any Class A
Bianchi
metric $h_{ij}$ such that ($\ga,\ h$) is an Einstein-Weyl structure, the Weyl
form $\ga$  may be written in our particular Gauduchon's gauge as :
$\ga = \Ga^ih_{ij}(T)\si^j$ , where the $\Ga^i$ are constant parameters.}

Indeed, at any given time $T$ one can find coordinates $\tilde{\si}^i$ such
that
$h_{ij}$ is a diagonal matrix $\tilde{h}_{ij}$, the structure constants being
unchanged. The matrix $\tilde{X}$ is then diagonal too, with elements
$$ \tilde{X}_{11} = [(n^{22}\tilde{h}_{22} -
n^{33}\tilde{h}_{33})^2]/(\tilde{h}_{22}\tilde{h}_{33})$$ and circular
permutations.
$h_{ij}$ being a strictly positive definite matrix, we get the vanishing of
$\frac{d\tilde{\ga}^i}{dT}\,,$  and, at that time of $\frac{d\ga^j}{dT}$ in any
coordinate frame ; the same results then holds at any proper time. Q.E.D.

We are now in position to discuss the issue of the diagonal hypothesis for the
metric $h_{ij}(T)\,.$

\noindent In the Einstein equation analysis, as explained by Tod \cite{Tod95},
the
condition (\ref{3-1-b}, with $\ga_i = 0$) ensures - at least for Bianchi $IX$
and
$VIII$ cases \footnote{\ As a matter of facts, Tod's argument can be also used
in
Bianchi $VI_{0}$ and $VII_{0}$ cases.} -, a possible
simultaneous diagonalisation of the matrices $h_{ij}$ and
$\frac{d h_{ij}}{dT}$ or
$K_{ij}$ at $T_0\,,$ with no change of the structure constants $n^{ij}\,.$

Here \footnote{\ We leave aside the Bianchi $I$ case, with vanishing structure
constants and matrix $X$, constant one-form coefficients $\Ga^i$ and where,
locally,
$\si^i \equiv dx^i$. Two of the $\Ga^i$ vanish, but one cannot prove that the
metric stay in a diagonal form.}, let us start from a proper time $T_0$ such
that
$h_{ij}$ (and $n^{ij}$) is diagonal. By inspection of the possible values of
$n^{ii}\,,$ equation (\ref{3-34}) ensures that the value of the constants
$\Ga^i$ fall into one of three cases :
\begin{itemize}
\item all zero : in particular, this is the sole solution in the Bianchi $VI_0$
case. In such a situation, there exists no non-exact Einstein-Weyl structure,
and Tod's argument ensures that for Bianchi $VI_0\,,\ VII_0\,,\ VIII\,$ and
$IX$ cases, there is no loss of generality in the choice of a diagonal metric
$h_{ij}(T)$ .
\item at most one of them vanishes : this may happen only in the Bianchi $IX$
case with $\tilde{h}_{ij}(T_0) =  h_0 \de_{ij}\,.$ Then, a possible
simultaneous
diagonalisation of the matrix
$\frac{d \tilde{h}_{ij}}{dT}$ or
$\tilde{K}_{ij}$ at $T_0\,$ is possible, and (\ref{3-34}), at $T=T_0 +
\epsilon$, enforces
the equalness of the $\tilde{\tilde{K}}_{ii}$ at $T_0\,.$ So, at that time, the
matrices $\tilde{\tilde{n}}\,,\ \tilde{\tilde{h}}\,,\ \tilde{\tilde{K}}$ are
proportional to the 3$\times$3 unit matrix. Then, one can find new coordinates
where
$\frac{d\tilde{\tilde{K}}}{dT}$ is also diagonal, which ensures that the
matrices
$h$ and $K$ stay in a diagonal form. But, equation (\ref{3-1-c}), where the
term
$\tilde{\tilde{\ga}}_i\tilde{\tilde{\ga}}_j$ is not in a diagonal form,
contradicts the hypothesis of at most one of the $\Ga^i$ vanishing.
\item one of them subsists : \begin{itemize}
\item this is the case for Bianchi $II$ case, but (\ref{3-1-b}) enforces no
further constraint on the metric and it seems hard to prove that the metric
will
stay in a diagonal form ;
\item this occurs in Bianchi $VII_0\,,\ VIII\,$ and $IX$ cases, when at that
time,
one of the
$\tilde{X}_{ii}$ given previously vanishes, say
$\tilde{X}_{33}$ .  For these three cases, we have, for a non-exact
Eintein-Weyl structure:

$$\tilde{n}^{11} = \tilde{n}^{22} = +1\ \ ,\ \ \tilde{\Ga}^1 = \tilde{\Ga}^2 =
0\
,\
\
\tilde{\Ga}^3
\neq 0\ ,\
\
\tilde{h}_{11}(T_0) = \tilde{h}_{22}(T_0)\,.$$ Condition (\ref{3-1-b}) ensures
that at $T_0$ :
$$\frac{d\tilde{h}_{31}}{dT} = \frac{d \tilde{h}_{32}}{dT} = 0\,.$$ As a
consequence, the particular block diagonal structure of the matrices
$\tilde{h}_{ij}\,,\ \tilde{n}^{ij}$ and
$\frac{d \tilde{h}_{ij}}{dT}$ ensures that they may be simultaneously
diagonalised
at
$T_0\,.$ So $\tilde{\tilde{h}}_{ij}$ and $\tilde{\tilde{K}}_{ij}$ (thanks to
equ.
 (\ref{3-1-c})) stay diagonal and we have proved that the constraints that
result
from Einstein-Weyl equations for Bianchi $IX$, $VIII$ and $VII_0$ in the
non-diagonal case, are the same as the ones in the diagonal situation.
\end{itemize}
\end{itemize}
\noindent We can summarize this discussion in a theorem :

\noindent {\bf Theorem 2} : {\it (Non-)compact Einstein-Weyl Bianchi metrics of
the types $VI_0,\ VII_0,\ VIII$ and $IX$ are conformally K\"ahler or
conformally
Einstein and the metric may be taken in a
diagonal form. In the non-exact Einstein-Weyl case, the metric and Weyl form
were given in equ.(\ref{3-47}). The conformal scalar curvature
has a constant sign on the manifold and, in our particular Gauduchon's gauge,
the
dual of the Weyl form is a Killing vector.}

Theorem 1 then gives the following :
\nl\noindent {\bf Corollary 2} : {\it (Non-)compact non-exact Einstein-Weyl
Bianchi IX metrics are conformally K\"ahler. The metric may be taken in a
diagonal form and is conformal to the 4-parameters' one given in
[GB.equ.(27)].}

\section {Concluding remarks} In this paper, we have presented a (nearly)
complete
analysis of the  Einstein-Weyl structures ($g\,,\
\ga$) corresponding to Class A Bianchi metrics. We have shown that, also in the
non-compact case, there exists a conformal gauge in which the conformal scalar
curvature is a constant, and we have proved that types $VI_0\,,\ VII\,,\ VIII$
and
$IX$, diagonal or not, are conformally K\"ahler or conformally Einstein.
We have explained why, in these cases, one can restrict oneself to
diagonal metrics. Moreover, in the non-exact Eintein-Weyl cases, the explicit
expression for the distance and Weyl 1-form, depending on 4
parameters submitted to some positivity requirements has also been obtained in
subsection 2.4.

The further requirement of completeness and compactness
will restrict the parameters of our solutions : in particular, Bianchi
$VI_{0},\ VII$ and $VIII$ metrics cannot give compact metrics, their isometry
group being non-compact. We shall give elsewhere the
full family of Compact Bianchi $IX$ Einstein-Weyl metrics, which, as we have
proven here, are conformally K\"ahler \cite{Bonneau98}.

Let us make a final comment on self-duality constraints on the Weyl connection
$\ga\,.$ In the vierbein basis corresponding to expression (\ref{3-47}), one
obtains
$$d\ga = \Ga^3((\Ga^3)^2 + u^2)\left[\frac{d}{du}[\frac{\Om^2}{-n^{33}u +
k}]e^0\wedge e^3 + n^{33}\frac{\Om^2}{(-n^{33}u + k)^2}e^1\wedge e^2\right]\
.$$ The (anti-)self duality of the Weyl connection then needs
$$\Om^2 = C(-n^{33}u + k)^{1\pm 1}\ .$$ Due to positivity requirements on
$\Om^2$, solutions exist only in the Bianchi
$IX$ case, and were given in [GB.Corollary 3]\cite{{PS93},{Madsen-b}}.

\section {Appendix}

Using equations (\ref{2-11}) and the definition of $K_i^j$ given in
(\ref{2-12}),
the Christoffel connection components are expressed as :
\beqa\label{A-0} 2\Ga^{\al}_{0\be} = \si^{\al}_i\si^j_{\be} K_j^i\ \ & , & \ \
2\Ga^{0}_{\al\be} = - \si_{\al}^i\si^j_{\be} K_{ij}\ \ ,\nnb \\
2\Ga^{\al}_{\be\ga} =  g^{\al\de}[g_{\de\be\,,\ga} + g_{\de\ga\,,\be} -
g_{\be\ga\,,\de}]\ \ & , & \ \ {\rm the\ other\ components\ vanishing}\ .
\eeqa Then, the covariant derivative of the three basis vectors
$\si^i_{\al}$ are found to be :
\beq\label{A-1}
\nabla_{\al}\si^i_{\be} = \partial_{\al}\si^i_{\be} -
\Ga^{(3)\ga}_{\al\be}\si^i_{\ga} = \frac{1}{2}C^i_{jk}\si^j_{\al}\si^k_{\be} +
h^{ij}h_{kl}C^k_{jn}\si^l_{(\al}\si^n_{\be)}\,.
\eeq The expression $$K_i^j\si_j^{\be}\nabla_{\be}\si^i_{\al} =
C^i_{jk}K_i^j\si^k_{\al}\ $$ will be useful, as well as
\beq\label{A-2}
\nabla_{\al}\si^{\al}_i = - \si^{\be}_i \si^{\al}_j\nabla_{\al}\si^j_{\be} =
C^j_{ij}\ .
\eeq The $\nu = 0$ component of the Bianchi identity
$2\nabla_{\mu}R^{(\nabla)\,\mu}_{\nu} = \nabla_{\nu}R^{(\nabla)}$ is split
according to $\mu = (0\,,\al)\ .$ Using (\ref{2-12},\ref{A-1}) and
$$R^{(\nabla)} = R^{(3)} + 2R^{(\nabla)}_{00} -\frac{1}{4}(\frac{h'}{h})^2 +
\frac{1}{4}K_{ij}K^{ij}$$ one obtains :
$$2\nabla_{\mu}R^{(\nabla)\,\mu}_{0} = \nabla_{0}R^{(\nabla)}
-h^{ij}\frac{dR^{(3)}_{ij}}{dT} + 2\nabla_{\al}^{(3)}R^{(\nabla)\,\al}_{0}\,.$$
As a consequence :
\beq\label{A-4} h^{ij}\frac{dR^{(3)}_{ij}}{dT} =
2[\nabla_{\al}^{(3)}\si^{\al}_k]R^{k}_{0}\,,
\eeq where $R^{k}_{0} = h^{ij}\si_i^{\al}R^{(\nabla)}_{0\al}\,.$

\bibliographystyle{plain}

\end{document}